\documentclass[manuscript,screen]{acmart}
\usepackage[utf8]{inputenc}
\usepackage{amsmath}
\usepackage{amssymb}
\usepackage{dsfont}
\usepackage{bbm}
\usepackage{subcaption}
\usepackage{multirow}
\setcopyright{none}

\usepackage[textsize=scriptsize,textwidth=2cm]{todonotes}

\definecolor{kentuckyblue}{RGB}{0, 93, 170}			

\newcommand{\users}{\mathcal{U}}
\newcommand{\items}{\mathcal{V}}
\newcommand{\features}{\mathcal{F}}
\newcommand{\domain}{\mathcal{D}}
\newcommand{\pro}{\mathcal{P}}
\newcommand{\sensitive}{\mathcal{S}}
\newcommand{\rerankers}{\mathcal{K}}
\newcommand{\metrics}{\mathcal{M}}

\newcommand{\reals}{\ensuremath{\mathbb{R}}}

\title{``And the Winner Is...'': Dynamic Lotteries for Multi-group Fairness-Aware Recommendation}

\author{Nasim Sonboli, Robin Burke}
\affiliation{%
  \institution{University of Colorado, Boulder}
  \city{Boulder}
  \state{CO}
  \country{USA}}
\email{ firstname.lastname@colorado.edu }

\author{Nicholas Mattei}
\affiliation{%
  \institution{Tulane University}
  \city{New Orleans}
  \state{LA}
  \country{USA}}
\email{  }

\author{Farzad Eskandanian}
\affiliation{%
  \institution{DePaul University}
  \city{Chicago}
  \state{IL}
  \country{USA}}
\email{feskanda@depaul.edu}

\author{Tian Gao}
\affiliation{%
  \institution{IBM Watson Research Center}
  \city{Yorktown Heights}
  \state{NY}
  \country{USA}}
\email{tgao@us.ibm.com}

\begin{document}

\begin{abstract}
As recommender systems are being designed and deployed for an increasing number of socially-consequential applications, it has become important to consider what properties of fairness these systems exhibit. There has been considerable research on recommendation fairness. However, we argue that the previous literature has been based on simple, uniform and often uni-dimensional notions of fairness assumptions that do not recognize the real-world complexities of fairness-aware applications. In this paper, we explicitly represent the design decisions that enter into the trade-off between accuracy and fairness across multiply-defined and intersecting protected groups, supporting multiple fairness metrics. The framework also allows the recommender to adjust its performance based on the historical view of recommendations that have been delivered over a time horizon, dynamically rebalancing between fairness concerns. Within this framework, we formulate lottery-based mechanisms for choosing between fairness concerns, and demonstrate their performance in two recommendation domains.
\end{abstract}

\maketitle

\section{Introduction}
In addition to the core property of accurate personalization -- delivering results that match user interests and preferences -- recommender systems may need to satisfy other, non-accuracy, constraints in certain applications. One property of interest that has received significant attention recently is \emph{fairness}: a constraint that a recommender system should try to distribute its benefits fairly across different stakeholder groups \cite{yao2017beyond,burke2018balanced,ekstrand2018exploring,liu2019personalized,kamishima2016model,beutel2019fairness}. For example, all else being equal, a job recommender system should not recommend executive jobs to male users and clerical jobs to female users.

Key stakeholder groups in recommender systems are often identified as \textit{consumers}, individuals who receive recommendations from the systems, \textit{providers}, stakeholders for the items that are being recommended, and the \textit{system} or platform. Fairness concerns may arise from either consumers or providers~\cite{burke_multisided_2017}. In the job recommendation case above, it was fairness across groups of consumers that is of interest. 

In this paper, we focus on the problem of \textit{provider fairness}: namely, how to ensure that a recommender system, over time, is recommending items from protected groups in a fair manner relative to others. We are interested in a multi-aspect version of this problem, where items may be associated with multiple, intersecting, protected groups.

Our motivating example is drawn from the peer-to-peer micro-lending platform, Kiva.org. The users of Kiva are lenders, who support entrepreneurs usually from developing countries by lending small amounts. The organization has the goal of providing equitable access to capital across different regions, economic sectors and borrower demographics. This organizational mission needs to be embedded in any system deployed to recommend loans to funders. Without some control over the characteristics of the recommendations delivered, it is easy to imagine that a positive feedback loop \cite{sun2019debiasing} could develop in which some types of loans are increasingly disadvantaged by the algorithm.

In general, we may anticipate that a fairness-aware recommender system will need to respond to multiple \textit{fairness concerns} simultaneously. In this work, we adopt a social choice perspective \cite{BCELP16a} on balancing different fairness concerns, which gives us a rich set of normative properties and algorithms. Social choice is fundamentally concerned with combining preferences from multiple parties into a single outcome in which all parties participate, voting being a paradigmatic example of such a choice~\cite{Zwicker:Voting}. We can think of each fairness concern as a kind of actor, with preferences over which recommendations should be delivered. Combining the preferences of multiple such concerns fits squarely into the social choice realm.

We believe that social choice is a more flexible and realistic framework for representing fairness-aware issues in machine learning than the optimization frameworks typically employed. Social choice is inherently multi-agent, and therefore, the idea of the integration of multiple fairness concerns naturally emerges, rather than being a complex add-on. Importantly for recommendation problems, social choice naturally allows for heterogeneity and hence personalization across decision instances since a user is just another agent with preferences over the outcome. Finally, fairness is inherently a social and political construct, and a social choice formalization allows the preferences of different actors to be foregrounded, rather than relegated to the black box of machine learning optimization. The study of fairness has a long history in the social choice literature \cite{Young:Equity,Zwicker:Voting}.

\paragraph{Contribution}
In this paper we propose a novel framework for recommender systems we call \textit{Social Choice for Re-ranking Under Fairness} (SCRUF). SCRUF uses multiple fairness metrics to evaluate the history of recommendation delivery and determine whether and how to adjust its performance. It uses feature-specific re-rankers to improve fairness, selecting one re-ranker (possibly non-deterministically) at each time point. We use a set of social choice inspired algorithms to allocate re-rankers to users based on the user preferences. This framework abstracts the particular fairness metrics away from the recommendation algorithm design, an approach that becomes unwieldy when attempting to incorporate multiple fairness concerns. It also supports a dynamic balance between the interplay between personalization and fairness and is therefore sensitive to context and individual differences. We demonstrate the efficacy of our design on two recommendation domains.

\subsection{Fairness-aware Recommendation}

A substantial body of research on fairness in machine learning, especially in classification settings, has emerged in the past ten years, including formalizing definitions of fairness~\cite{chouldechova2017fair,dwork2012fairness,hardt2016equality,narayanan2018translation} and offering algorithmic techniques to mitigate unfairness~\cite{kamiran2010discrimination,pedreshi2008discrimination,zemel2013learning,zhang2017anti}. Fairness in recommender systems emerged as a research topic more recently, first in the work of Kamishima et al. in 2012~\cite{kamishima2012enhancement}, and the topic has since drawn increased attention~\cite{burke_multisided_2017,yao2017beyond,kamishima2018recommendation,burke2018balanced,fatrec-workshop-2017,fatrec-workshop-2018,pmlr-v81-ekstrand18b,mehrotra2018towards}. Recommender systems, while a subclass of machine learning systems, are different enough that the results from classification cannot be readily applied. Chief among these challenges is the issue of personalization. A recommender system is supposed to deliver suggestions tailored to each user's preferences, providing every user with a different experience. As such, it differs from a classifier that establishes a classification function with a single decision boundary for all cases.

Both in the machine learning and the recommender systems formulation of fairness, there has been little recognition of the intersection of multiple fairness definitions and dimensions, although recent work has noted the benefits of combining multiple fairness definitions~\cite{beutel2019fairness}. Most existing research considers only a single protected class, and even in cases where multiple groups are considered as in~\cite{buolamwini2018gender,hebert2018multicalibration,kearns2017preventing,zhu2018fairness}, fairness is conceived the same way for all groups. In recognition of the complexity of the fairness concept, we seek to accommodate different definitions of fairness put forward by different stakeholders, all of which must be integrated in a single framework. This nuanced understanding of the value of fairness is essential for capturing the richness of this social construct in real-world settings such as those studied by scholars of organizational justice, anti-discrimination law, and social justice.

Two standard approaches have emerged to integrating fairness concerns into recommender systems. The \textit{integrated} approach builds a fairness constraint into the recommendation model itself, for example as a regularization constraint, balancing between accuracy and fairness in the optimization process for the recommendation model \cite{kamishima2012fairness,yao2017beyond}. The \textit{re-ranking} approach applies fairness to the output of a recommendation algorithm, reordering the results. Re-ranking approaches offer a number of advantages. First, the trade-off between accuracy and fairness can be tuned without re-learning the recommendation model. Second, researchers have found that re-ranking can achieve better trade-offs versus accuracy with this type of model~\cite{pmlr-v81-ekstrand18b,abdollahpouri2019managing,liu2019personalized}. Due to these advantages we choose to use the latter method.

There has been some recent work in recommendation fairness that incorporates multiple fairness dimensions, most notably \cite{sonboli-umap-2020}. This work integrates user tolerance towards different types of variation in item features with a representation of protected groups that spans multiple dimensions. The authors were able to show a beneficial trade-off between fairness and accuracy and improved results across different categories of protected groups. One drawback of the method of \cite{sonboli-umap-2020} is that it relies on weights associated with item features to boost the inclusion of protected group items into recommendation lists. Balancing across different groups requires careful setting of these weights and sometimes unexpected interactions arise. 

In addition, this and similar methods are list-wise approaches, which aim to increase protected group representation in individual lists. However, as noted above, the real objective of fairness-aware recommendation is to enhance fairness as measured historically, across the behavior of the system as it provides recommendations to many users over time. The personalization element of recommendation means that this objective cannot be targeted directly: any given user may or may not constitute a good opportunity to enhance fairness relative to a particular protected group. In \cite{sonboli-umap-2020}, this aspect of the problem was addressed by incorporating user-specific weighting, which can be interpreted as a preference in a social choice setting. 

Analyzing user characteristics enables the system to determine which users constitute good opportunities to pursue different fairness goals. However, there is another side of the problem. At any point in time, the system's historical performance may have been more favorable to one protected group than another. If we only look at the users, we ignore the signal from past performance about where fairness needs are most critical.

\subsection{Computational Social Choice for Fair Recommendations}

Fairness has been extensively studied in the economic field of social choice including work focusing on the division of continuous resources such as land or water \cite{Moulin:FairDivision}, on more discrete, indivisible settings such as goods and services \cite{Thomson:FairRules,Thomson:IntroFairAllocation} and more fundamentally in the areas of political economy having to do with justice and fair distribution of resources to individuals \cite{Young:Equity,Rawls:Justice,Rescher:Justice}. In its classical formulation, \emph{social choice} concerns itself with the study of how groups, where each member is endowed with their own preferences, make decisions that must be then shared by that group \cite{Sen:CollectiveChoice}. To these considerations the field of computational social choice adds computational tools including algorithms, complexity, and big data \cite{BCELP16a,DBLP:conf/ijcai/Mattei20}.

From the literature on social choice we will focus on the \emph{allocation} setting (which is a generalization of the classical matching setting) \cite{BCELP16a}. In allocation, the items within $A$ are to be distributed or \emph{allocated} to the set of agents in $N$.  Hence, the social part of social choice reinforces the idea that a set of preferences need to be considered and combined since the outcome of a social choice process will affect all the agents.  There have been many practical applications of matching and allocations from kidney allocation \cite{Roth:Kidney} to conference paper reviewing \cite{LiMaNoWa18}.  There are extensive studies of algorithms for a variety of settings \cite{Manlove:MatchingPrefs} and the study of fair allocations in multi-agent systems is a popular topic in the broad area of artificial intelligence \cite{Aziz:FairAllocation}. Equity and other concerns, formalized as economic axioms have a long history in social choice both in allocation \cite{Young:Equity} and voting \cite{Zwicker:Voting}. It is this long history of study of the axioms, or properties, of the algorithms and aspects including \emph{fairness} we hope to leverage.

Rather than thinking of integrating the concerns of protected groups into re-ranking decisions indirectly, in the form of weights for particular feature values as in \cite{sonboli-umap-2020}, our approach \textit{Social Choice for Re-ranking Under Fairness} (SCRUF) conceives of both users and protected groups as actors with preferences over the items that may be recommended. The goal is to achieve an integration of these preferences over the whole recommendation history. We explore a class of solutions to this problem that assumes multiple fairness criteria can be persued at the same time by deciding which objectives to address and which users to address them with.  We make this choice is made non-deterministically and also take into account both the current state of the recommendation history, which we can think of as defining immediate needs, and the user's propensity towards different item categories, which define current opportunities, in the context of providing personalized recommendation results. 

A social choice perspective on recommendation has emerged in recent research as a possible source of methods to integrate the viewpoints of multiple agents  or priorities \cite{burke2020algorithmic}. \citet{chakraborty2019equality} build a recommendation system for finding fair group recommendations through viewing them as elections between various signals of popularity, leading to a shared group recommendation.  This does not take the important aspect of \emph{personalization} into account that we address in SCRUF.  \citet{suhr2019two} explore driver assignment in two sided matching markets with an emphasis on producer and rider fairness. \citet{patro2020fairrec} propose a recommender system for two-sided matching markets with the goal of fair exposure amongst producers.  This differs from our work in that the fairness metrics are fixed and embedded into the matching algorithms themselves.  Finally, \citet{lee2019webuildai} propose a system that uses social choice to embed normative properties for algorithmic governance into the algorithms themselves, as demonstrated on a food bank matching scenario. However, none of this research considers multiple fairness concerns on the provider side of a recommendation system as required in the Kiva case or the dynamic response to historical fairness outcomes as embodied in SCRUF.

\section{Formal System Specification}

To leverage the power of recommender systems for personalized fairness we will define a set of choice functions to promote fairness. We first detail the formal notation for our recommender system and how to view it as a social choice problem. We then describe our overall system in terms of choice functions and how these can be used to promote fairness.

\subsection{Recommendation Systems with Protected Values}
In a recommendation system setting we have a set of users $\users = \{u_1, \ldots u_n\}$ and a set of items $\items = \{v_1, \ldots, v_m\}$. For each item $v_i \in \items$ we have a $k$-dimensional feature vector $\vec{v_i} = \langle f_{i1}, \ldots f_{ik} \rangle$ over a set of categorical features $\features = \{F_1, \ldots, F_k\}$, where each feature $F_i$ has finite domain $\domain_i$. We assume that all elements in $\items$ share the same set of features. 
Consider a running example of a funding site that shows micro-loans in emerging markets to potential funders. In this example we have $m=3$ items in the database which share $|\features|=4$ features: $\{$Region, Gender, Sector, Amount$\}$, where each has its own domain.  For example, $\domain_1 = \{$Africa, Middle-East, India$\}$.  This setting is illustrated in Table \ref{table:user_profile}.
    
\begin{table}
    \begin{tabular}{|c|c|c|c|c|}
    \hline
            & $F_{1}$ : Region & $F_{2}$ : Gender & $F_{3}$ : Sector & $F_{4}$ : Amount \\
    \hline
        $v_1$ & Africa & Male & Agriculture & \$0-\$500\\
    \hline
        $v_2$ & Africa & Female & Health & \$500-\$1,000\\
    \hline
        $v_3$ & Middle-East & Female & Clothing & \$0-\$500 \\
    \hline
    \end{tabular}
    \caption{Set of Potential Loans.}
    \label{table:user_profile}
\end{table}

Though our items are comprised of a set of features, we start with the view that not all features should be treated the same. We assume that there is a subset of the features $\sensitive \subseteq \features$ that are denoted as \textit{sensitive} and there is a subset of values for each such feature, i.e., $\pro_i \subseteq \domain_i$ that constitute the \emph{protected values} of the sensitive feature. That is, given all items in the recommendation system, we have a subset of sensitive features, each of which may contain protected values. Turning back to our running example, we may wish to target $\features_1=\{Region\}$ as a sensitive feature and the values $\pro_1=\{$Africa, India$\}$ as the protected values. We could designate sensitive features and protected values based on operational goals such as regions or genders that are funded less frequently.

For our recommender system we have a personalized ranking function $R(u_i, \items) \rightarrow \sigma_i(\items)$, which given user $u_i$ and set of items $\items$ produces a permutation, i.e., a ranking, over the set of items for that user, i.e. a recommendation. As a practical matter, the recommendation results will always contain a subset of the total set of items, typically the head (prefix) of the permutation $\sigma_i$ up to some cutoff number of items.

\subsection{Re-Ranking Functions}

In order to promote fairness, we assume that we are also given a set of re-ranking functions $\rerankers = \{\kappa_1, \ldots, \kappa_{|\sensitive|}\}$, which are a set of functions, one for each sensitive feature.  For feature $j$, the re-ranking function $\kappa_j(\sigma) \rightarrow \sigma^{\prime}$ will take a permutation $\sigma$ and produce a new permutation $\sigma'$ of the set of items that is more ``fair'' towards the particular protected feature values associated with $\sensitive_j$. In real applications, the final recommendation slate is a short list of the most preferred items from this final, re-ranked permutation. 

For our system we assume a common form of all re-ranking functions, where the permutation is achieved by sorting items based on a score, and the score is a linear combination of the score from the recommender system (the determiner of the original $\sigma$ ranking) and a score based on the presence of the protected feature, such that protected group items are moved up in the ranking list~\cite{adomavicius2009improving}. The scoring function $\rho$ for user $u$, an item $v$, and a sensitive feature $j$ is defined as follows:

\begin{equation}
    \rho(u, v, j) \buildrel\triangle\over= \lambda_j  (R(u,v) + (1 - \lambda_j) \mathds{1}_{\{v \in \sensitive_j\}}
\end{equation}


\noindent The indicator function $\mathds{1}_{\{v \in \sensitive_j\}}$ has the value 1 if the item $v$ has a protected value of sensitive feature $\sensitive_j$, and 0 otherwise. $\lambda_j$ is a feature-specific parameter that controls the trade-off between accuracy (as represented by the original $\sigma$ ranking) and fairness (as represented by the boost given to protected items). All of the items in the list $\sigma$ are re-scored using $\rho$, sorted in decreasing order, and truncated to produce the final $\sigma^{\prime}$ recommendation list.

\subsection{Metrics for Fair Recommendation}

There are a wide variety of metrics that have been proposed for measuring the fairness of a recommendation result or set of recommendation results.  In our setting we are not concerned with the fairness of a particular recommendation but rather the \emph{history of recommendations} the system has generated over within some time window.  Hence we track the prior history of recommendations lists that have been generated $\vec{L} = [\ell_1, \ldots, \ell_{t-1}]$ for the users (with a slight abuse of notation) $\vec{U} = [u_1, \ldots, u_{t-1} ]$ that have appeared to the system.  

Rather than commit to one particular metric in our system, we assume a family of functions $\metrics_j: \vec{L} \times \vec{U} \rightarrow \reals$, one for each sensitive feature $\sensitive_j$, mapping from a set of recommendation results $L$ and the set of users $U$ to whom each of those results have been delivered to a value indicating the degree of fairness in the total set of results. We assume that a higher $\metrics_j$ values indicate a fairer result. Without loss of generality, we assume that each metric has values in the range $[0,1]$.

We will assume that the re-ranking functions have a non-decreasing impact on their associated fairness metric. That is, given a recommendation result $\sigma$, $\metrics_j(\sigma, u) <=  \metrics_j(\kappa_j(\sigma), u)$.  Because of this property, we can interpret a fairness score as indicating the relative number of times we want to select the different re-ranking functions. If the metrics were all equal, then the different re-ranking functions would be equally desirable.

Note that the inclusion of $\vec{U}$ as an argument to the $M_j$ functions allows us to include a family of fairness metrics that are sensitive to the user's level of interest in items that vary on different feature dimensions. Each recommendation result is then evaluated relative to the user to whom it is delivered. For example, even if our recommendation history tells us we should be favoring loans in the textile sector, it may not be as valuable to recommend such loans to the agriculture-focused user, as opposed to a user that has proved to be more flexible in which sectors they support.

\subsection{User Preferences for Fairness}

To incorporate the social choice aspects of the problem, user preferences over both the overall set of items as well as preferences about the \emph{re-ranking functions themselves} need to be taken into account. In a traditional social choice setting we have a finite set of agents $N = \{1, \ldots, n\}$ and a finite set of alternatives $A = \{1, \ldots, m\}$. Each agent $i \in N$ has a preference $\succsim_i$ over the alternatives. Typically these preference are expressed as a binary relation (weak or linear order) over the set of alternatives $A$.  

While the user preferences are handled by the personalized ranking function $R(u_i, \cdot)$ we will also incorporate the preference over the fairness functions themselves. To this end, replacing the preferences $\succsim_i$ above, we assume that for each user we are also given a vector of real numbers, $\vec{\tau}_{u_i} = \{\tau_1, \ldots, \tau_k\}$ of length $k$, which indicates the tolerance (preference) of $u_i$ for variation relative to feature $F_k$.  We can then view our problem as one of allocating re-ranking functions to users based both on the their preferences and on the current fairness status.
 
\cite{eskan2017-personalized-diversity} introduced the concept of personalized diversity in collaborative filtering using a user-specific measure based on information entropy. High entropy in a categorical distribution of user profile represents high interest of user in diversity. Liu et al. \cite{liu2018personalizing,liu2019personalized} integrated this concept for the first time in recommendation re-ranking using a quantity $\tau_u$, a user-specific measure of interest in diversity.


\begin{equation}
\vec{\tau}_u(F_{j})\buildrel\triangle\over=-\sum_{f \in F_j} P(f|u)\log P(f|u),
\label{eq:tau}
\end{equation}

\noindent where $P(f|u)$ is computed as the fraction of items in the user's profile that have the feature value $f$. This can be interpreted as the user's likelihood of liking items with that value. The higher the entropy value is for a user on a feature, the higher their tolerance to see diversity within that feature. We assume that we can interpret this as a \emph{preference} in the social choice sense. In our running example, a user may be particularly dedicated to a particular economic sector, agriculture for example, and may only have supported loans in this sector in the past.  Hence, they would have low tolerance for variation in this feature. Note that since these map onto $\reals$ we could both interpret these as ordinal rankings: as a preference order for user $u_i$, $\succsim_i$, over the set of $f_j$; or as cardinal valuations.

\subsection{Overall Framework}

\begin{figure}[tb]
    \centering
    \includegraphics[width=5.5in]{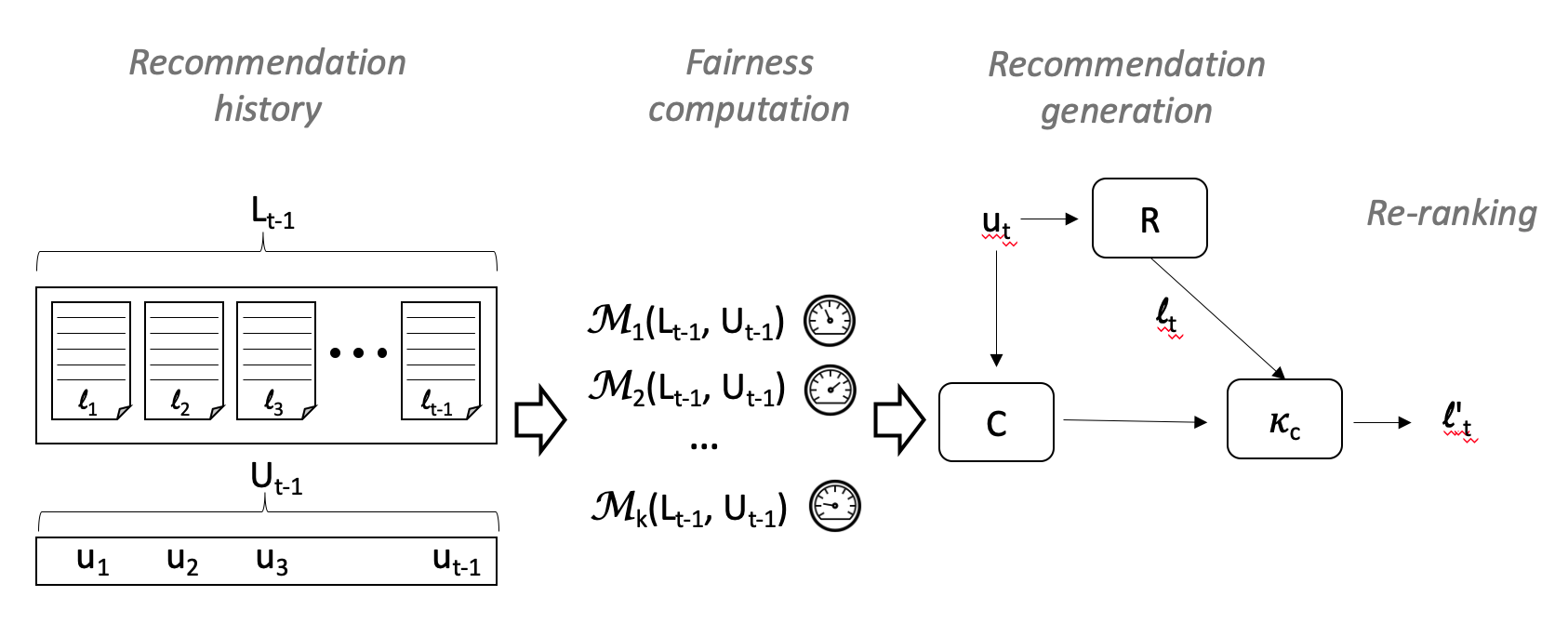}
    \caption{SCRUF framework, a snapshot at time $t$: On the left are the recommendation lists $L$ computed at prior time points. Fairness metrics $M$ compute the fairness state, which is input to the choice function $C$, selecting a re-ranker $\kappa_t$ that processes the recommendations $\ell$ from $R$ into a final re-ranked slate $\ell^{\prime}$.}
    \label{fig:framework}
\end{figure}

SCRUF is our framework for explicitly representing the design decisions that enter into trading off between accuracy and fairness across multiply-defined and intersecting protected groups in the setting described above. Figure~\ref{fig:framework} shows the general process that the framework instantiates by looking at a snapshot in time. A user $u_t$ arrives at the system and the base recommender algorithm $R(u_t,\items)$ generates a recommendation list $\ell_t$. 

SCRUF is able to accommodate different metrics, one for possibly each sensitive feature, $\features_j$, $\metrics_j: \vec{L} \times \vec{U} \rightarrow \reals$.  Since we have access to the history of all recommendations, we can derive particular fairness results relative to the different sensitive dimensions indicated by the meters associated with each metric $\metrics_j$. This set of metrics are input to a choice function $C$ which picks one dimension to prioritize and selects the corresponding re-ranking function $\kappa_c$, which is applied to $\ell_t$ resulting in a final set of recommendations $\ell^{\prime}_t$ that is displayed to the user. (As a practical matter, our implementation described below groups users into batches and calculates the fairness metrics only once per batch.)

Note the arrows from the user $u_t$ point both to the recommendation algorithm $R$ where the algorithm takes into account the user's inferred preferences over items in attempting to predict the user's preferences, and also to the choice function $C$ that may take into account the user's inferred preferences from their tolerance scores $\vec{\tau}_{u_i}$ over item features in choosing a re-ranking algorithm.

\subsection{Choice}
As noted above, we are investigating both deterministic and non-deterministic mechanisms for selecting, at each point when a recommendation is generated, a single $\kappa_j$ function to use to re-rank those recommendation to a specific user $u_j$. The choice function $C$ uses the current state of the recommendation history, as defined by the $M_j$ metrics over the recommendation history so far and optionally the identity of the current user, to compute a feature $c \in \sensitive$  whose corresponding re-ranking function $\kappa_c$ will be applied to the recommendations for this user, i.e., $\kappa_c(R(u_j, \items))$. We prefer this simple uni-dimensional re-ranking scheme over one that attempts to incorporate multiple fairness dimensions are considered at once because it creates independence between re-ranking operations and avoids complex parameter interactions that might occur in attempting to compute a single re-ranking incorporating multiple fairness dimensions.

One issue that may arise, and we discuss in the next section, is that we need to decide when to \emph{stop} running the re-rankers for a particular feature. If the historical data at time $t$ shows that we have been fair towards a particular feature, we do not need to promote it in this iteration. In the following we will describe how we select which features to consider.

\section{Choice Functions}

What remains to be specified within the SCRUF framework is the choice function $C$. There are wide variety of ways such a function could be formulated. In this work, we explore three different variants of the lottery, where probabilities are set for each re-ranker and a single re-ranker is chosen by sampling from this distribution.

In order to pick a choice function at time $t$ we will, for operational concerns, also have a parameter $w_b$ which defines the historical (backward) \emph{window} over which we are concerned with our fairness metric.  This means that our fairness metrics $\metrics_j$ are run over over the set of users and lists between $[t - w_b - 1, t-1]$.  Recall that our fairness metrics are all in the range $[0.0, 1.0]$ with $1.0$ representing full fairness for that metric.  In a slight abuse of notation we will use $\metrics$ to represent this list of values.

Also, for operational reasons, we are given an $\epsilon$ which represents a non-zero cutoff or tolerance for each metric.  We will only consider running re-rankers when $\metrics_j - \epsilon > 0.0$. This allows us to focus on the sensitive features with greater unfairness and provides a way to guard against an over-emphasize on protected groups at the expense of accuracy.

Let the \emph{unfairness vector} be (point-wise) $\vec{UF} = 1 - (\metrics - \epsilon)$.  Intuitively, this captures how unfair we are being towards a particular feature.  The following discussion will treat this vector as a probability distribution, and so it will be normalize to unit length, $\vec{UF} = UF / \sum_i(UF_i)$.  Note that this will implicitly assume that our target each element of $UF = 1/|UF|$, i.e., that we want unfairness to be equal across all aspects.  This is a byproduct of our metrics taking values in $[0.0,1.0]$ since if all metrics were $1.0$ then our normalization would be undefined, hence the $\epsilon$.  Normatively, this makes sense in that if we are being unfair the same amount to all sensitive features then we want an equal probability distribution over all features.

\subsection{Baselines}\label{sec:static}
We employ two simple baseline techniques to contrast with the more dynamic options discussed below. The simplest is the \textit{Fixed Lottery}, in which each re-ranker is chosen with equal probability for each set of recommendations delivered. This method has the benefit of great simplicity and does not require any bookkeeping about the historical fairness of the system. 

If we want to use the information in the $UF$ vector, another simple alternative is a deterministic \textit{Least Misery} algorithm, in which we identify the feature in the $UF$ with the highest value (most unfair) and chose the associated re-ranker. This method directs the system's attention to the dimensions with the worst historical performance and attempts to correct that. It will be dynamic in the sense that as the performance improves in one dimension, another may be chosen. 

\subsection{Dynamic Lottery}
We have found that it is typical for some dimensions to be more difficult to achieve fairness for than others. In particular, some types of items are rarely retrieved by the base recommendation algorithm and therefore only small improvements can be had through re-ranking. Applying the least misery algorithm in such a setting could lead the system to concentrate all of its effort on one of these intractable dimensions and miss opportunities to achieve fairness in other parts of the item space. 

To avoid this problem, we can use $UF$ as a lottery over the re-rankers and select a re-ranker with probability proportional to its weight, so that the poorest performing dimensions (most unfair) would have highest probabilities of being chosen. This is a \textit{Dynamic Lottery} as opposed to the fixed version above, because the probability associated with each re-ranker will change as a function of system performance. 

\subsection{Allocation Lottery}
While the above method is sensitive to the dynamic properties of the system, it is not sensitive to each user's particular propensity or interest towards different dimensions. In prior work, the ability to re-rank selectively based on user characteristics was found to yield a better tradeoff between accuracy and fairness~\cite{liu2019personalized,sonboli-umap-2020}. For this reason we consider in this section \emph{randomized allocation mechanisms} that consider both users and fairness concerns.  In a such a mechanism we compute a fractional allocation that we can then sample from in order to compute an assignment.  So, for a given set of $n$ agents and $m$ objects, we compute a bi-stochastic matrix of size $n \times m$ which represents the fraction of a particular item is allocated to an agent.

We use a modification of the \emph{probabilistic serial (PS)} mechanism \cite{bogomolnaia2001new}.  In PS, also known as the simultaneous eating algorithm, each object is considered to have an infinitely divisible probability weight of one.  To find the allocation every agent, simultaneously and at the same speed, begins ``eating'' their most preferred object that has not been completely consumed already.  Once an object is consumed, the agents move to their next most preferred object until all objects have been consumed. The random allocation of an agent by PS is the amount of each object he has eaten. PS satisfies a number of important fairness and efficiency criteria \cite{Aziz:EqulibriaPS,Aziz:EgalRandom} and has been used in real allocation settings such as course selection at universities \cite{budish2013designing}.

In translating our recommendation system setting to use PS we use again the tolerance values $\tau$ of the agents as representative of their preferences.  As PS only requires ordinal preferences we simply use the ordering and not the actual values (breaking ties randomly when needed). A key concept in PS is the idea of an object's capacity, how much it is available to be allocated. We set the capacity of each re-ranker to mirror the sampling lottery probability from above. Specifically, each re-ranked has weight $w_f * UF_i$, thus limiting the amount of that re-ranker to allocate.  We then run the PS algorithm and get a fractional allocation for each user for each re-ranker.  We interpret this fractional allocation (normalized into a distribution) as the probability that the user should be assigned that particular re-ranker.

\section{Methodology}
\subsection{Evaluation Metrics}
As our work here concentrates on ranking performance, we use normalized discounted cumulative gain (nDCG) as our measure of recommendation accuracy. Note that we are only evaluating re-ranking algorithms so nDCG is limited to some extent by the performance of the base algorithm to which the re-ranking is applied. 

Provider-side fairness metrics come in two basic varieties. There are those that respond to the appearance of protected items in a recommendation list: \textit{exposure} metrics, and those that take into account the suitability of the target user as \textit{hit-based} metrics~\cite{abdollahpouri-umuai-2020}. In this work, we concentrate on exposure metrics, in particular, protected class exposure, which calculates the fraction of a retrieved recommendation list belongs to a particular protected class. This value is related to the fairness concept of ``statistical parity,'' measured relative to items' level of promotion within the recommender system. Because list lengths are fixed (10 in our case), the exposure of unprotected items is just one minus the protected group exposure. We note, however, that exposure metrics may overstate the effectiveness of re-ranking, since they do not evaluate the quality of the protected items promoted into the recommendation list. Exposure $e_j$ of the protected class items relative to feature $S_j$ is defined as:

\begin{equation}
    e_j(\ell) = \frac{\sum_{v \in \ell}{\mathds{1}_{\{v \in \sensitive_j\}}}}{|\ell|}
\end{equation}

Given this definition, our fairness metrics use the notion of absolute unfairness \cite{yao2017beyond}, and have the following form:

\begin{equation}
    M_j(L, U) = 1 - | 1 - 2 \frac{\sum_{\ell^{\prime} \in L}{e_j(\ell^{\prime})}}{|L|} |
\end{equation}
where $L$ is the list of recommendation $L = [\ell_{t-w_b},...,\ell_{t-1}] $. Note that this definition implies ideal fairness consists of equal exposure, that is, recommendation lists containing 50\% protected group items.\footnote{The metric as defined penalizes lists with more than 50\% protected items, which might seem counterproductive. However, as a practical matter in our experiments higher exposure values for protected items were never achieved.} We plan to explore other characterizations of exposure and other fairness metrics in future work. 

Each metric has a maximum fairness of 1 and therefore it is possible to calculate regret $\omega_j$ as the difference between this ideal $\metrics_j^*$ and the current state of the metric $\metrics_j$. For reasons of space, we report only on the average regret over all metrics, and leave more detailed analysis for future work. To understand the consistency of algorithm performance, we also compute the variance of the average regret across time periods.

\subsection{Dataset}
We tested our model on two datasets. The first is The Movies Dataset, which was obtained from the Kaggle website and contains the metadata of 45,000 movies listed in the Full MovieLens Dataset \footnote{https://grouplens.org/datasets/movielens} which were released on or before July 2017. Although movies are not a domain to which important fairness concerns are typically applied, we use this dataset as a well-known example with a rich set of provider-side features. Additionally, we extracted two features that contain demographic information on the movie directors and screenplay writers.

The dataset contains 26 million ratings from 270,000 users for all 45,000 movies. Ratings are on a scale of 1-5. Each movie contains a set of features from which the following were used in this project: genres, original language, release date, run-time, popularity, director gender and writer gender. A sample of this dataset was extracted which contained the 361,468 ratings from 6,000 users on 6,037 items (density of 0.99\%). 

All the features with numerical values were transformed into categorical values. Release date is bucketed into four groups, run-time into six groups and popularity is bucketed into five groups. In this dataset, three types of genders were present: 0, 1, 2. And each movie can be directed or written by a group of directors or writers. To capture this diversity, gender was discretized into seven groups. For example if a movie is directed by all the genders, we assign 012 for the gender information and if it is directed only by one gender, a single number was assigned to that movie e.g. 0, 1 or 2. All the categorical features were transformed into dummy variables, resulting in a total of 335 binary features. Table~\ref{table:sensitive_features_table} shows some examples of the sensitive features and their protected values.

In a fielded application, the choice of sensitive features and protected groups within those features may be determined by legal liability or business model considerations. Lacking this type of insight, we chose to identify protected features as those associated with rarely-recommended items. To determine the protected values of each feature, we performed a trial run of recommendation generation over the data set, and examined the distribution of features in the results. In a live system, historical recommendation data would be available over which to calculate this distribution. The values in the 25th 
percentile of the distribution were selected as the protected group for that feature.

\begin{table}
    \begin{tabular}{|c|c|c|c|}
    \hline
        Dataset & Features & Protected Values & Unprotected Values \\
    \hline
        \multirow{3}{*}{Kiva} & 
        Activity & Bicycle Repair, Gardening, Souvenir Sales & Taxi, Fishing, Vehicle Repairs \\ 
        & Country & Indonesia, Nigeria, Yemen & Cameroon, Armenia, Lebanon \\
        & Gender & Male & Female \\
        
    \hline
        \multirow{3}{*}{MovieLens} & Genres & Documentary, Foreign, War, Western & Adventure, Crime, Action, Comedy\\
        & Writer Gender & \{'01', '012'\} & \{'0', '02', '12', '2', '1'\} \\
        & Director Gender & \{'01', '1'\} & \{'0', '12', '012', '02', '2'\} \\
    \hline
    \end{tabular}
    \caption{Examples of sensitive features and their values.}
    \label{table:sensitive_features_table}
\end{table}

Our algorithm is also evaluated on a proprietary dataset obtained from Kiva.org, including all lending transactions over an 12-month period. Initially, there were 1,084,521 transactions involving 122,464 loans and 207,875 Kiva users. Of these loans, we found that 116,650 were funded, that is they received their full funding amount from Kiva users by the 30-day deadline imposed by the site. We selected only the funded loans for analysis. Each loan is specified by features including borrower's name/id, gender, borrower's country, loan purpose, funded date, posted date, loan amount, loan sector, and geographical coordinates. To reduce the feature space, and to solve the multicollinearity problem, highly correlated features were removed. 

The percentage funding rate (PFR) was added as a new feature, computed as follows:
\begin{equation}
 \mbox{\textit{PFR}} =  \frac{1}{\mbox{\textit{{\#} days to fund}}} * 100 
\end{equation}

The percentage funding rate captures the speed at which a loan goes from being introduced in the system to being fully funded.\footnote{Loans not fully funded within 30 days are dropped from the system and the money raised is returned to lenders.} For example, a loan with PFR of 25\% is accumulating a quarter of its needed capital each day. After preparing the data, the final features for each loan reduced to borrower's gender, borrower's country, loan purpose, loan amount (binned to 10 equal-sized buckets), and loan's percentage funding rate. We found that this dataset was highly sparse (density = $4.2e^{-5}$) and could not support effective collaborative recommendation, because a loan can only attract a limited amount of support (up to that needed for its funding). There are no ``blockbuster'' loans with thousands of lenders.

To generate a denser dataset with greater potential for user profile overlap, we applied a content-based technique creating \textit{pseudo-items} that represent groups of items with shared features. We applied agglomerative hierarchical clustering \cite{rokach2005clustering} using the features of borrower gender, borrower country, loan purpose, loan amount (binned to 10 equal-sized buckets), and percentage funding rate (4 equal-sized buckets). We chose the cluster with the highest Silhouette Coefficient \cite{rousseeuw1987silhouettes} of around 0.69 which indicates a reasonable cohesion of the clusters. Then we applied a 10-core transformation, selecting pseudo-items with at least 10 lenders who had funded at least 10 pseudo-items. The retained dataset has 2,673 pseudo-items, 4,005 lenders and 110,371 ratings / lending actions.

To identify the protected values for each feature, we applied the same method as for the MovieLens data set. We assigned the values that their frequencies are in the 25 percentile of the distribution to the protected group for each feature. The final number of features are 231 for this dataset.

\subsection{Experiments}
Our experimental methodology is designed to highlight differences between these choice functions. We followed a typical recommendation evaluation process with each user's profile split into 80\% training and 20\% testing. We chose non-negative matrix factorizing (NMF) as our base algorithm~\cite{takacs2008investigation} based on prior experience with these data sets. We plan to explore the interaction between base algorithm and choice functions in future work. The factorization model was built using the training data and then used to generate a recommendation list $\ell$ for each user.
Arrival time was simulated in our experiments. Users were shuffled randomly and grouped into batches of size 0.5\% of all the users, where each batch was considered to be a single time step. For each batch, we computed fairness metrics $\metrics$ over the previous 20 batches, so that the backward window $w_b$ equals approximately 10\% of the test data. \footnote{For the first batch, when no backward window exists, the Fixed Lottery was performed.} The experiment was run for each of the four choice functions described above: Fixed Lottery, Deterministic Least Misery, Dynamic Lottery, and Allocation Lottery. For the choice functions dependent on $\metrics$, we computed the lottery probabilities once per batch. 

The results of the different algorithms were compared in summary and over the course of each experiment's iterations. Overall nDCG was compared to establish the accuracy loss for each choice function. Over the course of each experiment, we computed cumulative fairness regret on each fairness dimension and on average. 

\section{Results}

\subsection{Overall results}

Table~\ref{tab:overall-ML} shows the overall results for the MovieLens data set. The first point to notice is that fairness is greatly improved (5x) over the base algorithm for all of the re-ranking methods, which is to be expected. Interestingly, the Fixed choice function, which chooses among the re-rankers with equal probability has the best fairness over all experiment iterations taken as a whole. The other re-rankers are similar. All of the re-rankers show a reduction in ranking accuracy, around 25\% of nDCG. We did not seek to minimize nDCG loss in these experiments as doing so would reduce the impact of any given re-ranking operation and require a longer experiment to tease out differences.

\begin{table}[htb]
\setlength\tabcolsep{0pt}
    \begin{subtable}{.5\textwidth}
    \centering
    \begin{tabular}{c|l|l|l}
       Algorithm  & \ nDCG \ & \ Fairness \  & \ Fairness Variance \ \\
       \hline
       Base (NMF)       & \ 0.143 & \ 0.039 & \ 5.3e-6 \  \\
       Fixed            & \ 0.107 & \ 0.179 & \ 3.8e-3 \ \\
       Least Misery \   & \ 0.106 & \ 0.178 & \ 1.3e-3 \ \\
       Dynamic          & \ 0.104 & \ 0.170 & \ 1.5e-3 \ \\
       Allocation       & \ 0.109 & \ 0.171 & \ 2.3e-4 \ \\
    \end{tabular}
    \caption{MovieLens data set}
    \label{tab:overall-ML}
 \end{subtable}%
   \begin{subtable}{.5\textwidth}
    \centering
    \begin{tabular}{c|l|l|l}
       Algorithm  \ & \ nDCG \ & \ Fairness \ & \ Fairness Variance \ \\
       \hline
       Base (NMF)   \ & \ 0.057 & \ 0.214 & \ 2e-4\\
       Fixed        \ & \ 0.045 & \ 0.323 & \ 1.1e-3\\
       Least Misery \ & \ 0.043 & \ 0.322 & \ 9e-4\\
       Dynamic      \ & \ 0.045 & \ 0.325 & \ 1e-3\\
       Allocation   \ & \ 0.048 & \ 0.327 & \ 1e-4\\
    \end{tabular}
    \caption{Kiva data set}
    \label{tab:overall-Kiva}
    \end{subtable}
    \caption{Summary results. Fairness measured by percentage of protected item exposure in recommendation lists.}
\end{table}

\begin{figure}[tbh]
\setlength\tabcolsep{0pt}
    \begin{subfigure}{1.0\textwidth}
    \centering
    \includegraphics[width=4.5in]{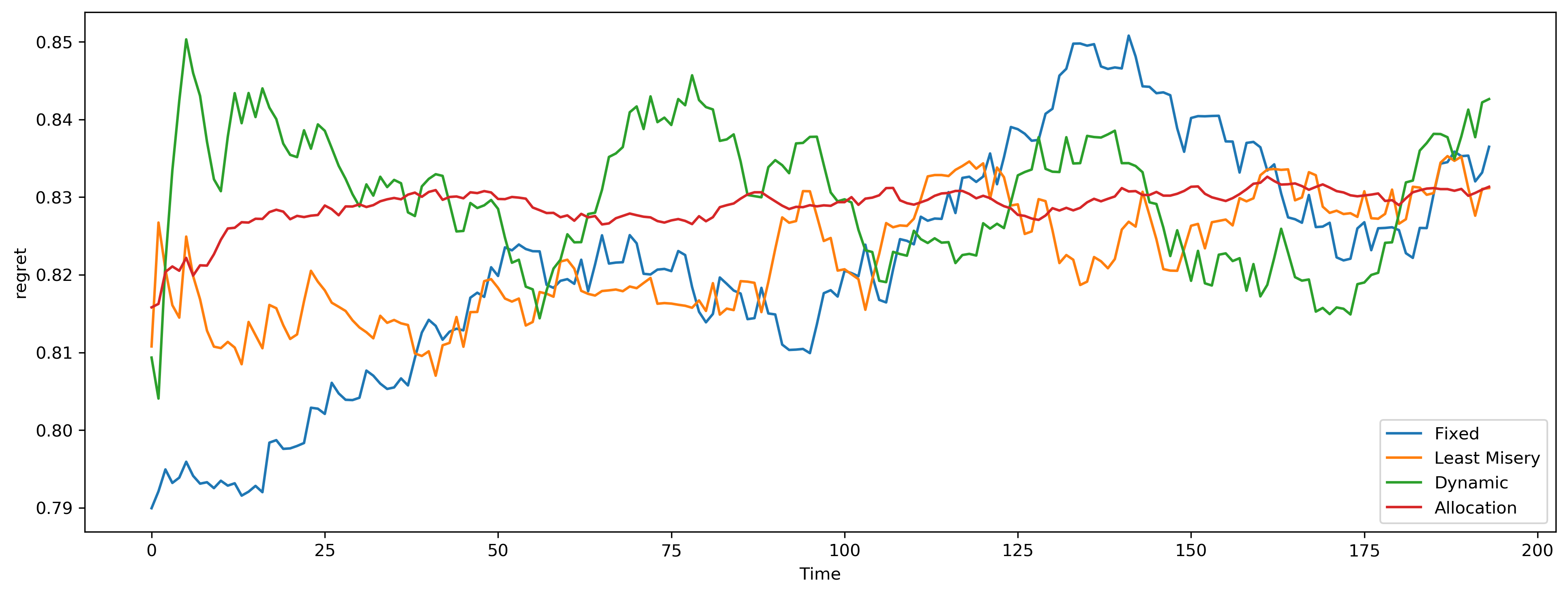}
    \caption{MovieLens data set}
    \label{fig:local-regret-ML}
    \end{subfigure}
    \begin{subfigure}{1.0\textwidth}
    \centering
      \includegraphics[width=4.5in]{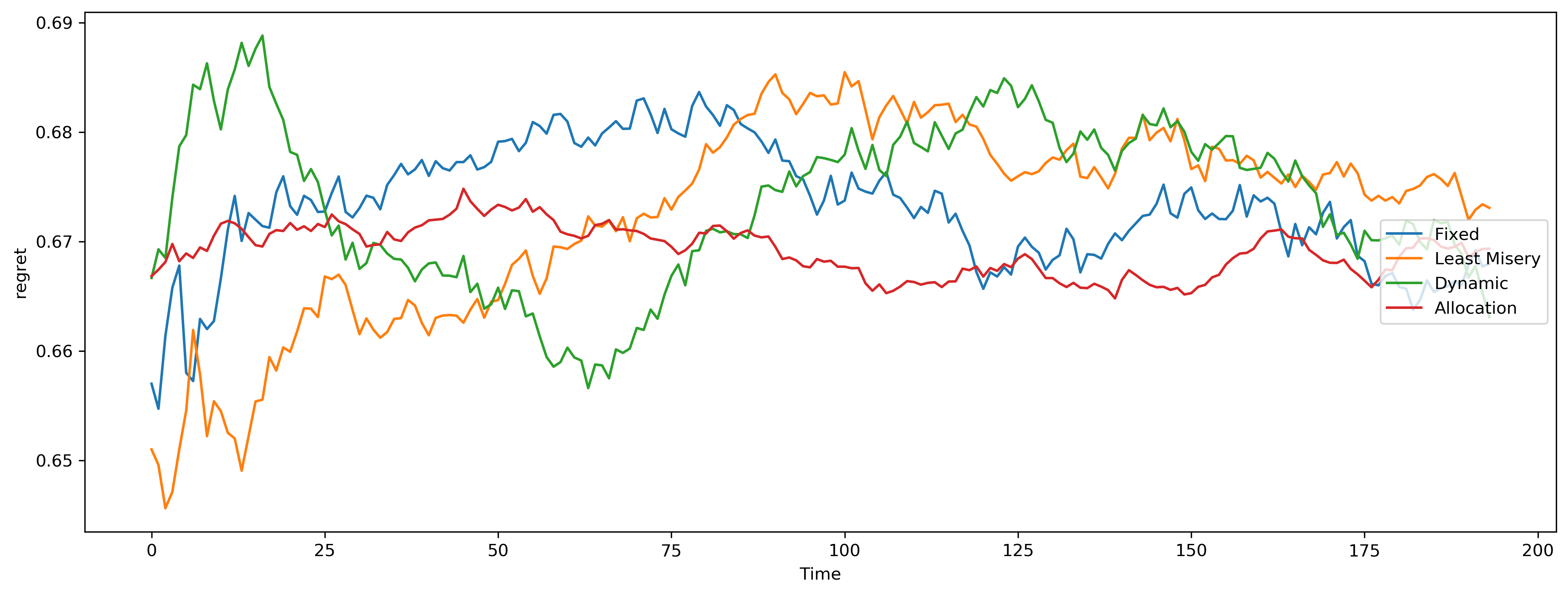}
    \caption{Kiva data set}
    \label{fig:local-regret-Kiva}
    \end{subfigure}
    \caption{Average fairness regret over time}
    \label{fig:local-regret}
\end{figure}

Table~\ref{tab:overall-Kiva} shows similar results for the Kiva data set. Here we do not see as much accuracy loss. The Allocation algorithm, which here has the best nDCG, is only 5.5\% below the original base algorithm. For this data set, the re-rankers also improve fairness, although not as dramatically as in the MovieLens case. The Allocation method has the highest fairness score in addition to the best nDCG.
Figure~\ref{fig:local-regret} shows the average fairness regret over time for the experiment. The algorithms all move within a fairly narrow regret bound, indicating the difficulty of achieving fairness in these data sets. The Fixed lottery shows lowest regret over most epochs for the MovieLens data set, but does not do as well with Kiva. Similar inconsistency is shown with the Least Misery algorithm. The low variance of the fairness of Allocation algorithm can be seen, as its regret does not show the swings of the other algorithms.

\begin{figure*}[t!]
    \centering
    \begin{subfigure}[t]{0.5\textwidth}
        \centering
        \includegraphics[width=3.0in]{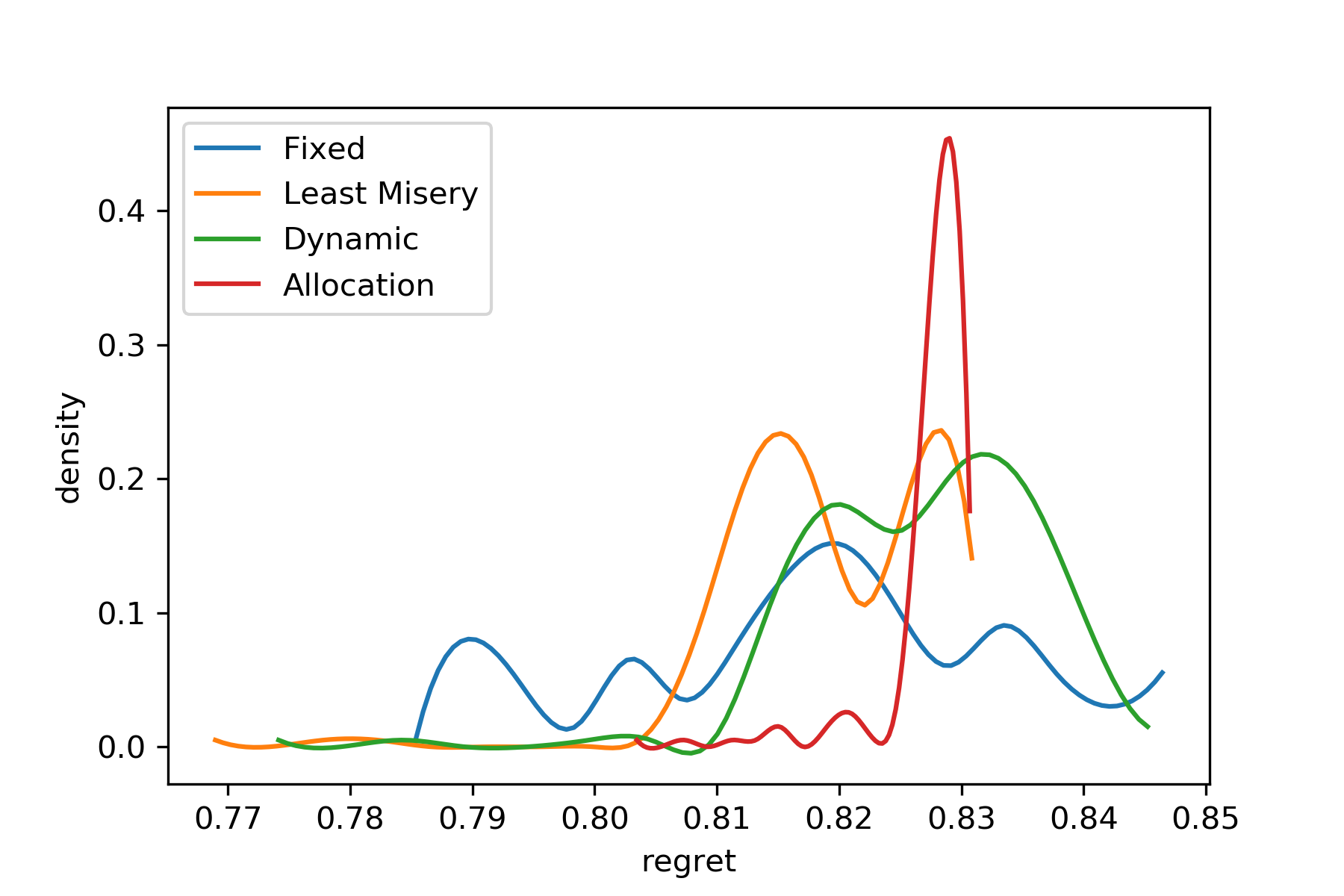}
        \caption{MovieLens}
        \label{fig:local-regret-ML}
    \end{subfigure}%
    ~ 
    \begin{subfigure}[t]{0.5\textwidth}
        \centering
        \includegraphics[width=3.0in]{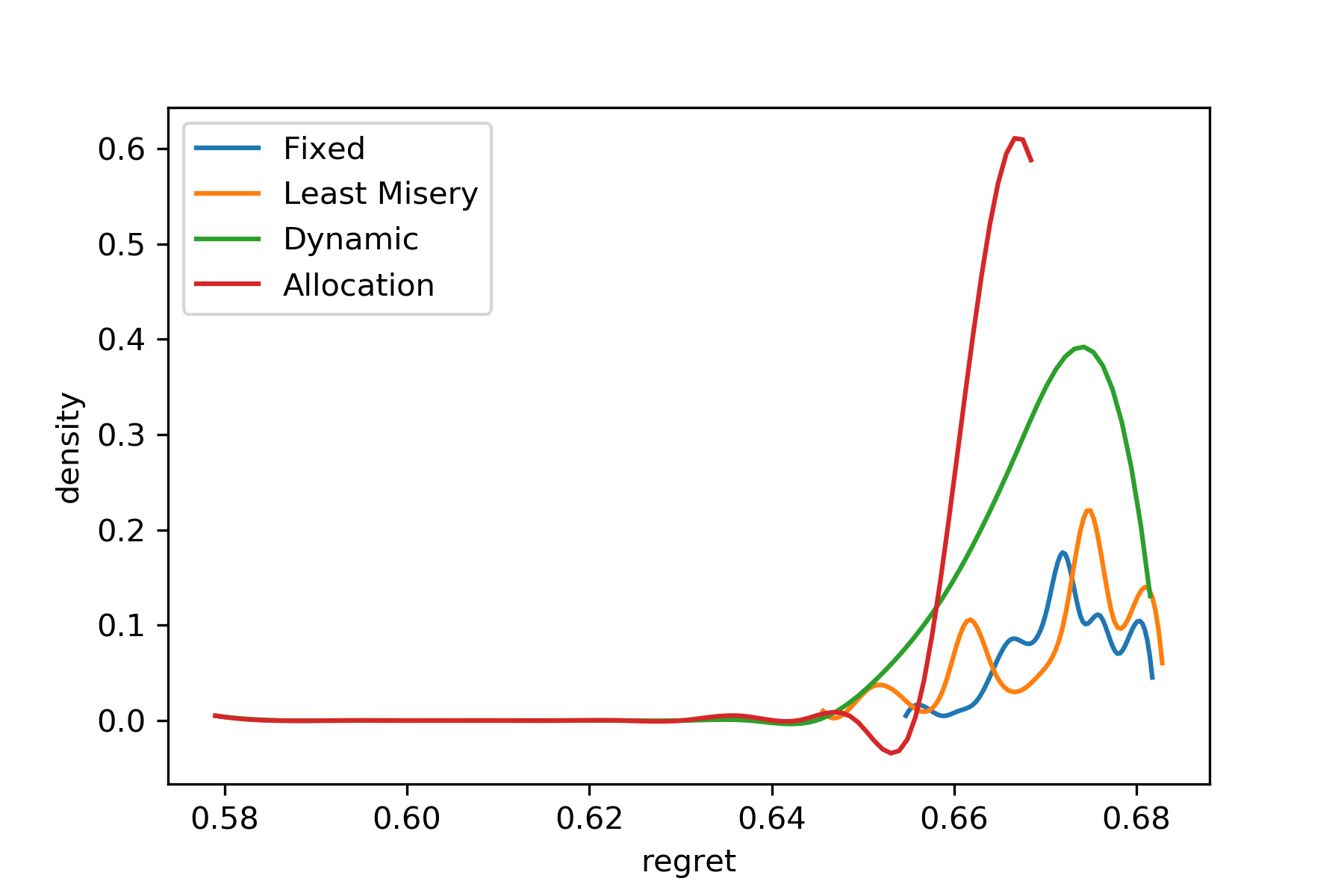}
        \caption{Kiva}
        \label{fig:local-regret-Kiva}
    \end{subfigure}
    \caption{Distribution of fairness regret}
\end{figure*}

Across both tables and in the time series figure, we see that the Allocation method has  lower variance in the fairness it achieves across iterations. Another way to see this consistency is the distribution of the average regret values. Figures~\ref{fig:local-regret-ML} and \ref{fig:local-regret-Kiva} show the distribution of regret for the different choice functions on each data set. The distribution of the Allocation method (shown in red) falls within a much narrower band than any of the other methods, particularly in the MovieLens data set where we see the Fixed method in blue taking on a wide range of regret values. At any given time, the Allocation algorithm is producing consistently fair results without the large variations in regret seen the other algorithms. As its fairness results are similar to those of the other lottery mechanisms, this consistency is a good reason to prefer it.

\section{Conclusion}
In this paper, we conceptualize algorithmic fairness and recommendation fairness, in particular, as a problem of \textit{social choice}. That is, we define the task of computing a recommendation as a problem of arbitrating among the preferences of different individual agents to arrive at a single outcome. For our purposes, the agents in question include the user and also multiple \textit{fairness concerns} that may be active within a particular organization. 

The move to frame fairness as a problem of social choice has several important consequence. First, it highlights the multiplicity and diversity of fairness (and other stakeholder) concerns that might be relevant in a given application. This approach allows us to be agnostic to different definitions and metrics of fairness and does not impose any particular structure on stakeholder preferences.

Second, we are able to make use of the large body of research in computational social choice, including the study of fairness, that has emerged in the past decades. 

Building on these ideas, we demonstrate the SCRUF framework for dynamic adaptation of recommendation fairness using social choice to arbitrate between different re-ranking methods. We define a set of choice functions, ranging from a simple fixed lottery to an adaptation of the probabilistic serial mechanism, and demonstrate their performance on two data sets where multiple fairness concerns have been defined. We found relatively minor differences between the different lottery mechanisms, except that the Allocation mechanism, which takes user preferences over features into account, provides lower variance in fairness over time and therefore a more consistently fair output.

\section{Acknowledgements}
Authors Burke and Sonboli were supported by the National Science Foundation under Grant No. 1911025.

\bibliographystyle{ACM-Reference-Format}
\bibliography{allocation.bib,recsys.bib}
\end{document}